\documentclass[aps,prl,twocolumn,superscriptaddress,amsmath,amssymb]{revtex4}
\usepackage{graphicx} 
\usepackage{epstopdf} 
\usepackage{dcolumn} 
\usepackage{xcolor} 
\usepackage{amsmath,amssymb} 
\usepackage{hyperref} 
\usepackage[utf8]{inputenc} 
\usepackage[english]{babel} 
\usepackage{ifthen} 
\usepackage{orcidlink} 
\usepackage{etoolbox}
\usepackage{physics} 

\graphicspath{{figures/}} 

\usepackage{multirow} 

\newboolean{articletitles}
\setboolean{articletitles}{true} 

\input{belle2-symbols}

\newboolean{wordcount}
\setboolean{wordcount}{false} 

\ifthenelse{\boolean{wordcount}}%
{\usepackage{bibentry} 
 \usepackage{comment} 
 \excludecomment{align*} 
 \excludecomment{align} 
 \excludecomment{equation*} 
 \excludecomment{equation} 
 \excludecomment{eqnarray*} 
 \excludecomment{eqnarray} 
 \excludecomment{acknowledgments} 
 \def\maketitle{} 
}{}

\begin{document}


\title{Search for \BdToKsttautau decays at the Belle II experiment
}

\author{I.~Adachi\,\orcidlink{0000-0003-2287-0173}} 
\author{K.~Adamczyk\,\orcidlink{0000-0001-6208-0876}} 
\author{L.~Aggarwal\,\orcidlink{0000-0002-0909-7537}} 
\author{H.~Ahmed\,\orcidlink{0000-0003-3976-7498}} 
\author{H.~Aihara\,\orcidlink{0000-0002-1907-5964}} 
\author{N.~Akopov\,\orcidlink{0000-0002-4425-2096}} 
\author{M.~Alhakami\,\orcidlink{0000-0002-2234-8628}} 
\author{A.~Aloisio\,\orcidlink{0000-0002-3883-6693}} 
\author{N.~Althubiti\,\orcidlink{0000-0003-1513-0409}} 
\author{M.~Angelsmark\,\orcidlink{0000-0003-4745-1020}} 
% ...

\collaboration{The Belle II Collaboration}

\begin{abstract}
We present a search for the rare flavor-changing neutral-current decay \BdToKsttautau with data collected by the \belletwo experiment at the SuperKEKB electron-positron collider. 
The analysis uses a \lumiyfours data sample recorded at the center-of-mass energy of the \FourS resonance. 
One of the \B mesons produced in the $\FourS\to\Bz\Bzb$ process is fully reconstructed in a hadronic decay mode, while its companion \B meson is required to decay into a \Kstarz and two \mtau leptons of opposite charge. 
The \mtau leptons are reconstructed in final states with a single electron, muon, charged pion or charged \Prho meson, and additional neutrinos. 
We set an upper limit on the \Branching of $\BR(\BdToKsttautau) < \ULdata$ at the 90\% confidence level, which is the most stringent constraint reported to date. 
\end{abstract}

\maketitle

In the standard model (SM) $\bquark\to\squark \tau \tau$ transitions occur only at the loop level via penguin or box diagrams, and are therefore suppressed. 
Combining the charged and neutral \BToKsttautau modes~\cite{KstDef}, the predicted SM \Branching is $(0.98 \pm 0.10) \times 10^{-7}~$\cite{Hewett1996, Capdevila2018}.
New physics models can enhance $\bquark\to\squark \tau \tau$ rates by up to four orders of magnitude~\cite{Capdevila2018}. 
In some scenarios, the leading new physics couplings are those involving the third-fermion generation, making $\bquark\to\squark \tau \tau$ transitions a better probe than $\bquark\to\squark e e$ and $\bquark\to\squark \mu \mu$~\cite{Cornella:2021sby}. 
Enhancements to $\bquark\to\squark \tau \tau$ are also foreseen in models that explain the recently observed 2.7$\sigma$ departure from the SM expectation in the \BtoKnunu decay rate~\cite{Belle-II:2023esi, Allwicher:2023xba}, together with the $b\to c\tau \nu$ anomalies~\cite{banerjee:2024, Capdevila2018}. 

The \belle experiment has reported an upper limit at the 90\% confidence level on the \BdToKsttautau \Branching of $3.1 \times 10^{-3}$, using a 711~\invfb data sample recorded at the \FourS resonance and reconstructing the accompanying \B meson in fully hadronic decay modes~\cite{Dong2023}. 
Other $\bquark\to\squark \tau \tau$ searches have been conducted by \babar~\cite{Lees2017} and \lhcb~\cite{Aaij2017}. 
None of these searches led to evidence for a signal, and all the upper limits are above the SM predictions. 

We report a search for \BdToKsttautau ~\cite{chargeConj} decay at \belletwo. 
The main challenge is the presence of up to four final-state neutrinos.
Reconstructing one \B meson (\Btag) in $\epem\to\FourS\to\BBbar$ isolates the accompanying \B meson (\Bsig) and constrains the kinematics of the missing neutrinos. 
We reconstruct the \Btag in hadronic decay modes and search for a \BdToKsttautau decay of the partner \Bsig, where the $\Kstarz\to K^+\pi^-$ mode is used. 
All the combinations of the decay modes $\tau^- \to e^- \overline{\nu}_e \nu_{\tau}$, $\tau^- \to \mu^- \overline{\nu}_{\mu} \nu_{\tau}$, $\tau^- \to \pi^- \nu_{\tau}$, and $\tau^- \to \rho^- \nu_{\tau}$ are utilized for the two \Ptau's. 
A multivariate approach with a binary classifier is adopted to separate signal from background events. 
The classifier output is used in a binned profile-likelihood fit to extract the signal \Branching. 
To minimize experimental bias, we finalized the full analysis procedure before examining the data in the signal region. 
The main new features compared to the previous \BdToKsttautau search~\cite{Dong2023} are the use of an improved tagging method~\cite{Keck:2018lcd} and the inclusion of $\tau^- \to \rho^- \nu_{\tau}$ decay modes. 
The \belletwo analysis described here improves the expected sensitivity by a factor of 2.6, despite using a smaller dataset. 

We use a \lumiyfours data sample recorded by the \belletwo detector, located at the SuperKEKB asymmetric electron-positron collider~\cite{Akai:2018mbz}, between 2019 and 2022. 
The data are collected at a center-of-mass (\CM) energy $\sqs = 10.58\gev$, corresponding to the peak of the \FourS resonance. 
An additional \lumioff data sample, collected at a \CM energy 60\mev below the mass of the \FourS resonance (off-resonance), is used to study backgrounds from \epem\to\qqbar events, where \quark indicates a \uquark, \dquark, \squark, or \cquark quark. 
The \belletwo detector~\cite{Abe:2010gxa} consists of a nearly hermetic magnetic spectrometer, composed by silicon detectors and a central drift chamber (CDC), surrounded by particle identification (PID), electromagnetic calorimetry (ECL), and muon and \KL subdetectors. 
Simulated samples are used to suppress backgrounds, estimate the signal efficiency, and define fit templates for the \Branching extraction. 
Various Monte Carlo (MC) event generators are utilized. 
The \texttt{EvtGen}~\cite{Lange:2001uf}, \texttt{PYTHIA8}~\cite{Sjostrand:2014zea}, \texttt{KKMC}~\cite{Jadach:1999vf}, and \texttt{TAUOLA}~\cite{Jadach:1990mz} software packages are used to model particle production and decay, \texttt{PHOTOS}~\cite{Barberio:1993qi} is used for photon radiation from final state charged particles, and \texttt{Geant4}~\cite{Agostinelli:2002hh} simulates material interaction and detector response. 
We simulate \BdToKsttautau signal decays using SM form factor calculations from Ref.~\cite{Ali:1999mm}. 
Simulated beam-induced backgrounds are overlaid onto the events~\cite{Natochii:2022vcs}. 
The \belletwo analysis software framework~\cite{Kuhr:2018lps, basf2-zenodo} is used for event reconstruction. 

The analysis starts by reconstructing a \Btag into one of 32 hadronic \Bz decay modes using the Full Event Interpretation algorithm (FEI)~\cite{Keck:2018lcd}, in which final state particles, reconstructed from tracks and energy deposits in the ECL (clusters), are combined into intermediate particles until the final \Btag candidates are formed. 
For each decay chain, the algorithm calculates the probability (\pFEI) that the reconstructed decay mode is correct, using gradient-boosted decision trees~\cite{FastBDTBelleII}. 
Known \Btag kinematic properties can be exploited using the beam-energy-constrained mass $\Mbc = \sqrt{ s/(2c^2)^2 - \left| \vec{p}^{\ast}_{\Btag} / c \right|^2 }$ and the energy difference $\deltaE = E^{\ast}_{\Btag} - \sqs / 2$, where $\vec{p}^{\ast}_{\Btag}$ and $E^{\ast}_{\Btag}$ represent the momentum and energy of the \Btag in the \CM frame, respectively. 
Correctly reconstructed $B$ mesons peak in the \Mbc distribution at the known \Bz mass and have values close to zero for $\Delta E$. 
We select \Btag candidates with $\Mbc > 5.27\gevcc$, $\left| \deltaE \right| < 200\mev$, and $\pFEI > 0.01$. 
At this stage, for the simulated signal sample, 0.9\% of events are retained, with an average multiplicity of 1.5 \Btag candidates. 
To minimize possible mismodeling in the $\Btag-\Bsig$ combination step, for each event only the \Btag candidate with the highest value of \pFEI is retained. 
In the signal MC sample, this selection picks the correctly reconstructed \Btag candidate about 87\% of the time. 

Tracks and clusters not associated to the selected \Btag are used to reconstruct \Bsig candidates. 
Spurious tracks from beam-induced background are rejected by selecting only the tracks that originate close to the interaction point by requiring small transverse and longitudinal impact parameters, $dr < 2 \cm$ and $|dz| < 4\cm$, respectively. 
Furthermore, tracks should have polar angles in the CDC acceptance ($17^{\circ} < \theta < 150^{\circ}$) and transverse momenta $p_T > 100~\mevc$. 
In order to have a reliable d$E$/d$x$ measurement for PID, only tracks having at least 20 CDC hits are retained. 
For each track, a charged particle hypothesis is assigned using PID selectors, which combine information from the different sub-detectors. 
A likelihood-based binary classifier separates charged pions from kaons. 
For muon identification, a likelihood-based selector is used as well, while for electrons a selector based on a boosted decision tree (BDT)~\cite{Hocker:1019880} is applied. 
We choose selection thresholds resulting in efficiencies of 95\% for hadrons, 82\% for electrons, and 77\% for muons; in the selected sample, 6\%, 3\%, and 30\% of the respective candidates are misidentified, according to the \BdToKsttautau simulation. 
PID selection efficiencies and misidentification probabilities are evaluated via data control channels in bins of momentum and polar angle of the reconstructed particles~\cite{Belle-II:2023esi, Belle2:2020}. 
Simulated events are weighted to correct for the differences compared to the data. 
Clusters that are not matched to any extrapolated track, and have polar angles in the CDC acceptance, are used to identify photon candidates. 
We reconstruct \piz candidates using photons with energy thresholds of $80\mev$, $30\mev$, and  $60\mev$ in the forward, barrel, and backward region of the ECL respectively. 
These thresholds are optimized to suppress background from high multiplicity $B$ decays. 
Photon pairs are then combined if their invariant mass is in the range $(120,~145)~\mevcc$, corresponding to approximately two times the experimental resolution around the known \piz mass~\cite{ParticleDataGroup:2024cfk}, and their maximum angular separation is 1.5~\rad in the transverse plane and 1.4~\rad in the 3D space. 
The selection criteria result in a \piz reconstruction efficiency of about 30\%. 
The \piz reconstruction efficiency is computed in bins of \piz momentum using data control samples from $D$ and $\tau$ decays. 
The observed differences between data and simulation are used to correct the simulated distributions. 
The \piz candidates are then used to construct $\rho^- \to \pi^- \pi^0$ candidates. 
We require an invariant mass in the range (0.65, 0.9)~\gevcc, which is chosen to reduce the contamination from combinatorial background. 
Pairs of charged kaons and pions having opposite charges are combined to form $K^{\ast 0} \to K^+\pi^-$ candidates if their invariant mass is in the range (0.80,\, 0.99)~\gevcc, corresponding to four times the \Kstarz decay width. 
A vertex fit~\cite{Krohn2020} constrains the two tracks to originate from a common vertex, which is identified as the \Kstarz candidate decay point. 
Candidate \Kstarz mesons that have a \pvalue for the vertex fit smaller than $10^{-3}$ are rejected. 

Each \Bsig candidate is formed by combining a reconstructed \Kstarz with the daughters of two oppositely charged $\tau$ leptons, in combinations of \en, \mun, \pim, and \rhom. 
The daughters of the two $\tau$ leptons are henceforth denoted as $t_1$ and $t_2$, where $t_1$ has the same charge sign as the charged kaon from the \Kstarz. 
A total of 16 possible combinations of $t_1 t_2$ final states are examined. 
The invariant mass of the \Kstarz, $t_1$, and $t_2$ system should be less than 6.0~\gevcc, to reject badly misreconstructed combinations. 

Multiple \Bsig candidates can be reconstructed for a single event. 
At this stage, all the \Bsig candidates are retained and combined with the unique \Btag, to form \FourS candidates. 
For each \FourS candidate, the rest of the event (ROE) contains all the remaining tracks and clusters not used either for \Btag or for \Bsig reconstruction. 
We require zero charged tracks in the ROE. 
The track selection in the ROE is the same as for the signal candidate except that the requirement on the number of CDC hits is removed. 
The selection of clusters in the ROE also follows the procedure adopted for \Bsig photons, with energy thresholds of 100\mev, 60\mev, and 150\mev in the forward, barrel, and backward regions of the ECL, respectively, optimized to reject clusters from beam background. 
We also require no track trajectory extrapolating to the ECL in a 20\cm radius around the cluster center. 
The latter requirement suppresses background clusters due to secondary particles generated in nuclear interactions of hadrons with the ECL material. 
We define the extra energy (\Eecl) as the sum of the energies of all the clusters assigned to the ROE. 

To reduce the contamination from events with particles outside of the acceptance of the tracking detectors, the polar angle of the missing momentum should be inside the CDC angular acceptance. 
The missing four-momentum of the event is defined as $p_{\rm miss} = p_{\rm init} - \sum_i p_{i}$, where $p_{\rm init}$ is the initial total four-momentum of the colliding beams, and the sum runs over the four-momenta of all the tracks and all the clusters in the ECL without an associated track. 
The flavor of the reconstructed \Bsig, determined by the charge of the kaon from the \Kstarz decay, is required to be opposite to the flavor of the partner \Btag candidate. 
The opposite correlation is also used to define the ``same-flavor'' control sample, with reconstructed \BzBz and \BzbBzb pairs. This sample also includes signal events in which one of the two $B^0$ mesons has undergone flavor mixing before decaying, with a mixing probability $\chi_d = 18.6\%$~\cite{ParticleDataGroup:2024cfk}, resulting in a signal contamination below the expected sensitivity of the analysis. 
The \BdToKsttautau signal efficiency at this stage is $4.9 \times 10^{-4}$ with an expected background yield of $98.4 \times 10^{3}$ events, about 35\% of which are $\epem\to\FourS\to\BzBzb$ events. 
The next largest source of background, about 30\% of the total, is from $\epem\to\ccbar$ continuum processes. 

The kinematic features of the signal side depend on the decay modes of the two \Ptau leptons. 
Therefore, we separate the reconstructed events into four signal categories: three $t_{\mathrm{1}}t_{\mathrm{2}}$ categories, namely $\ell\ell$, $\pi\ell$, and $\pi\pi$, where {$t_{\mathrm{1}},t_{\mathrm{2}}$}={$\ell,\pi$} indicates the \Ptau daughter, and the $\rho$ category, in which one \Ptau decays as $\tau^-\to\rho^-\nu_{\tau}$, while the other \Ptau decays to any of the four reconstructed modes. 
The first three categories differ in the number of neutrinos in the final state. 
The $\rho$ category can have one or two reconstructed \piz's from \Bsig, while the number of neutrinos is not fixed. 
In the simulated signal sample, the average \FourS candidate multiplicity per event is 4.4, and the fraction of events with one correctly reconstructed \Bsig candidate is 77\%. 
To avoid double counting due to cross-feed among signal categories, we select a single \FourS candidate in each event, as follows. 
First, we select the \FourS candidate(s) having the closest \Kstarz mass to the measured value~\cite{ParticleDataGroup:2024cfk}. 
If a reconstructed $\rho^-\to \pi^- \pi^0$ is present in the event, we retain only the candidates assigned to the $\rho$ category to prevent the $\pi^0$ from contributing to the ROE. 
Candidates with more identified electrons or muons are then selected to maintain a high signal efficiency for the purest categories. 
At this stage, the average \FourS candidate multiplicity in each event is 1.3, and for the events that have multiple candidates, the final choice is performed randomly. 
A comparison using a fully random selection of the \FourS candidate shows that the procedure outlined above results in a 15\% higher fraction of correctly reconstructed signal events, with no bias introduced in the observables used for the signal extraction. 

\begin{figure}[tp]
    \centering
    \includegraphics[width = 3.in]{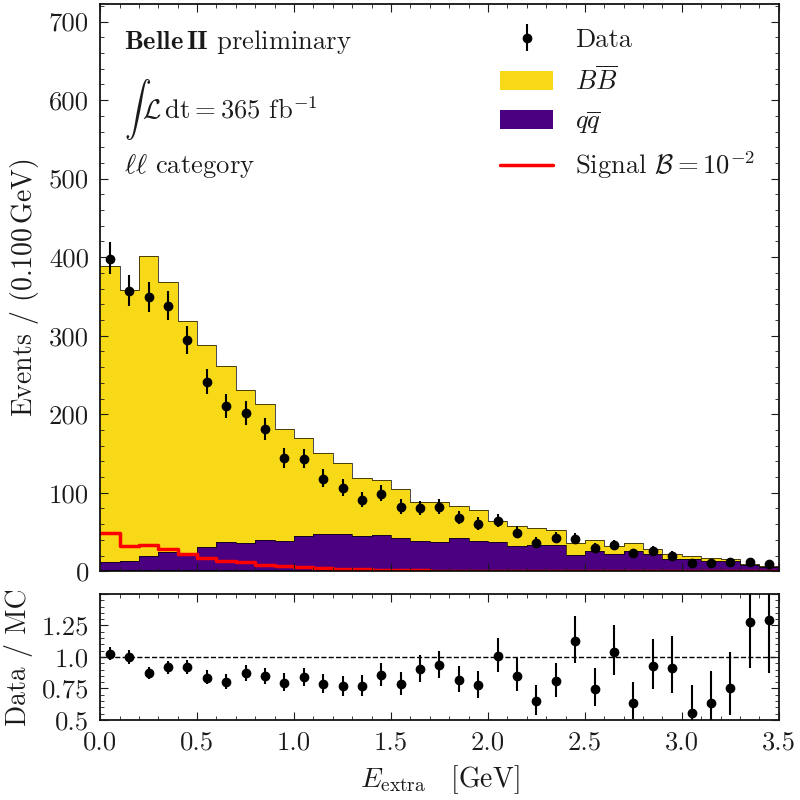}
    \caption{Distribution of \Eecl for events passing the nominal selection, with all the corrections applied, for the $\ell\ell$ signal category. The signal \BdToKsttautau histogram is shown scaled assuming a \Branching of $10^{-2}$. }
    \label{fig:distr_eecl}
\end{figure}

A control sample with a clean $\BJpsiKstar(\tomumu)$ decay is used to validate the simulated \BdToKsttautau signal, using the same strategy described in~\cite{Belle-II:2023esi}. 
The tracks and clusters associated with the $\jpsi\Kstarz$ candidate are replaced with those originating from a $\Kstarz\tautau$ decay, extracted from simulated \BdToKsttautau events, so that only the \Btag and the ROE's are taken from the control sample.
A difference in efficiency between data and simulation of $0.81 \pm 0.09$ is found and the central value is used as a correction for the \BdToKsttautau signal efficiency. 
The same factor is also used to scale the \BzBzb background events having a correctly reconstructed \Btag (peaking \BzBzb). 
The remaining portion of the \BBbar events (combinatorial \BBbar) is normalized with the same-flavor control sample. 
The same-flavor sample is also used to correct the ROE cluster multiplicity, separately for each signal category, since MC simulation mismodelling affects the \Eecl distribution. 
Figure~\ref{fig:distr_eecl} shows the \Eecl distribution for the $\ell\ell$ category with the ROE cluster multiplicity correction applied. 
The residual data/MC discrepancies for $\Eecl > 200$ \mev are covered by the systematic uncertainties, described below. 
The normalization of \epem\to\qqbar backgrounds events is adjusted for each signal category using the off-resonance data sample. 

We separate signal from background through BDT binary classifiers~\cite{Chen:2016:XST:2939672.2939785}, trained on simulated samples, separately for each signal category. 
Each BDT uses the same set of 14 variables, selected based on their discriminating power and on the level of data-simulation agreement observed in the same-flavor control sample. 
These variables combine information about the event shape, the kinematics of \Kstarz and $\tau$ candidates, the missing four-momentum, and \Eecl. 
In addition, the invariant masses $M(K^{\ast0};\,t_i)$, $i=1,2$ are used as inputs to target $\B \to \D^{(*)}\ell\nu$ backgrounds. 
The most discriminating input variables are the missing energy, \Eecl, $M(K^{\ast0};\,t_i)$, and $q^2 = (p_{\tau^+} + p_{\tau^-})^2$~\cite{supplemental}. 
To use the entire simulated sample, we perform a two-fold training: the samples are randomly split into two halves, and the classifier is trained separately on each half. 
The set of model weights obtained from each training are then applied to the complementary half sample. 
Good agreement between the outputs of the two trainings is observed. 

The range $[0.4, 1]$ of the BDT output ($\BDT$), defined as signal region (SR), is used in the fit for the \Branching. 
The SR is common to all signal categories and is determined by the need to maintain high signal efficiency while limiting the impact of background-related systematic uncertainties on the expected \Branching. 
Table~\ref{table:yields_signal_region} gives the signal efficiencies ($\varepsilon$) and the expected background composition in the SR. 
The $\ell\ell$ category is the most sensitive. 
In this category, the largest background contamination comes from \BBbar, 60\% of which are peaking \BzBzb pairs. 
In about $45\%$ of the \BBbar events, a neutral \Btag is correctly reconstructed while the other \Bz decays to a semileptonic or semitauonic final state. 
This fraction is lower for the other signal categories and is $20\%$ for the $\pi\pi$ one. 

\begin{table}
    \caption{Signal efficiencies ($\varepsilon$) and expected background yields, for $\BDT > 0.4$. The signal categories are ordered according to the expected sensitivity.}
    \label{table:yields_signal_region}
    \begin{ruledtabular}
    \begin{tabular}{crrr}
	Signal category &  $\varepsilon \times 10^{5}$ & \BBbar  & \qqbar    \\
	\hline
	$\ell\ell$   & $4.0$ & $275$	& $39$	\\
	$\pi\ell$    & $7.6$ & $1058$	& $230$	\\
    $\rho$       & $15.5$ & $3279$	& $845$	\\
	$\pi\pi$     & $4.0$ & $1077$	& $424$	\\
    \end{tabular}
    \end{ruledtabular}
\end{table}

The signal \Branching is extracted from an binned maximum likelihood fit to the \BDT distribution in the SR, simultaneously for all four signal categories, using the \textsc{pyhf}~\cite{pyhf, pyhf_joss} and the \textsc{Cabinetry} libraries~\cite{Cranmer2021}. 
The parameter of interest is $\BR(\BdToKsttautau) = N_{\rm sig} / ( 2 \varepsilon N_{\FourS} \fzz )$, where $N_{\FourS} = (387 \pm 6)\times 10^6$ is the number of produced \FourS, estimated from a data-driven approach in which non-\FourS events are subtracted from on-resonance data, and $\fzz=0.4861^{+0.0074}_{-0.0080} $~\cite{banerjee:2024} is the $\FourS\to\BzBzb$ decay rate. 
A uniform bin width is chosen for each signal category, such that each bin contains more than 10 expected background events. 
The templates for the fit components (signal, \BBbar and \qqbar backgrounds) are obtained from simulated samples. 

The parameter of interest is unconstrained in the fit, while systematic uncertainties are incorporated into the likelihood as nuisance parameters with Gaussian constraints. 
The uncertainties on the correction factors for the pion, kaon, and lepton identification efficiencies and misidentification probabilities, as well as for the \piz efficiency, are determined from the auxiliary measurements, as described above. 
We assign a systematic uncertainty to the correction of the ROE cluster multiplicity. 
The uncertainty is taken to be 100\% of the residual difference in the data-to-simulation ratio observed in the $\BDT<0.4$ control region, after applying the correction derived from the same-flavor sample. 
The \Branchings of decay modes contributing to about 70\% of \Bz decays and 50\% of \Bp decays in the SR are allowed to vary according to their known uncertainties~\cite{ParticleDataGroup:2024cfk}. 
We assign a 50\% uncertainty on the \Branchings of $B\to \D^{\ast\ast}\ell/\tau \nu$ decays, which are poorly known and constitute about 5\% (9\%) of the residual \BzBzb (\BpBm) background. 
The \BBbar events in which a $D$ meson decays to a final state with a \KL are scaled by 1.30, and a 10\% systematic uncertainty is assigned to them~\cite{Belle-II:2023esi}. 
The normalization factors for \qqbar, combinatorial \BBbar, peaking \BzBzb backgrounds, and signal efficiency are evaluated in the SR using the same control samples previously described, and they are found to be consistent with the values determined in the full \BDT region. 
Thus no further correction is applied; systematic uncertainties associated with the corrections are described below. 
An uncertainty on the normalization factor for the \qqbar background, due to the statistical uncertainty of the off-resonance sample and ranging from 70\% for the $\ell\ell$ mode to 15\% for the $\rho$ and $\pi\pi$ signal categories, is assigned. 
The combinatorial \BBbar yield is allowed to vary by $15\%$, while the peaking \BzBzb and signal normalizations are assumed to be known at the 14\% level. 
We use a dedicated \BdToKsttautau MC sample, generated with modified form factors~\cite{Ali:1999mm}, to evaluate the systematic uncertainty due to the knowledge of signal form factors. 
Global normalization uncertainties on the luminosity measurement (0.5\%), the number of \FourS (1.5\%), and the \fzz parameter ($^{+1.5\%}_{-1.6\%}$) are treated with one nuisance parameter each. 
Finally, the systematic uncertainty due to the limited size of simulated samples is also taken into account. 

Before examining the SR, we validate the fit procedure with MC pseudo-experiments, in which both statistical and systematic uncertainties are taken into account. 
No bias in the \Branching and its uncertainty is observed, with an injected signal \Branching ranging from zero to the current upper limit value~\cite{Dong2023}. 
As an additional check, a set of pseudo-experiments is constructed by varying the number of expected events in each bin of the fit variable. 
The variations are derived from the data-simulation discrepancies observed in those bins for the same-flavor control sample. 
Again in this case, no bias is observed when performing a fit. 
Assuming the background-only hypothesis, the expected \Branching uncertainty on simulated events is computed to be \ERRexpected. 
This corresponds to an expected 90\% confidence level (C.L.) upper limit of \ULexpected, which was determined using the CLs method~\cite{Read2002}, a modified frequentist approach that is based on a profile likelihood ratio~\cite{Cowan2011}. 

\begin{figure*}[htp]
\centering
\includegraphics[width = 7.in]{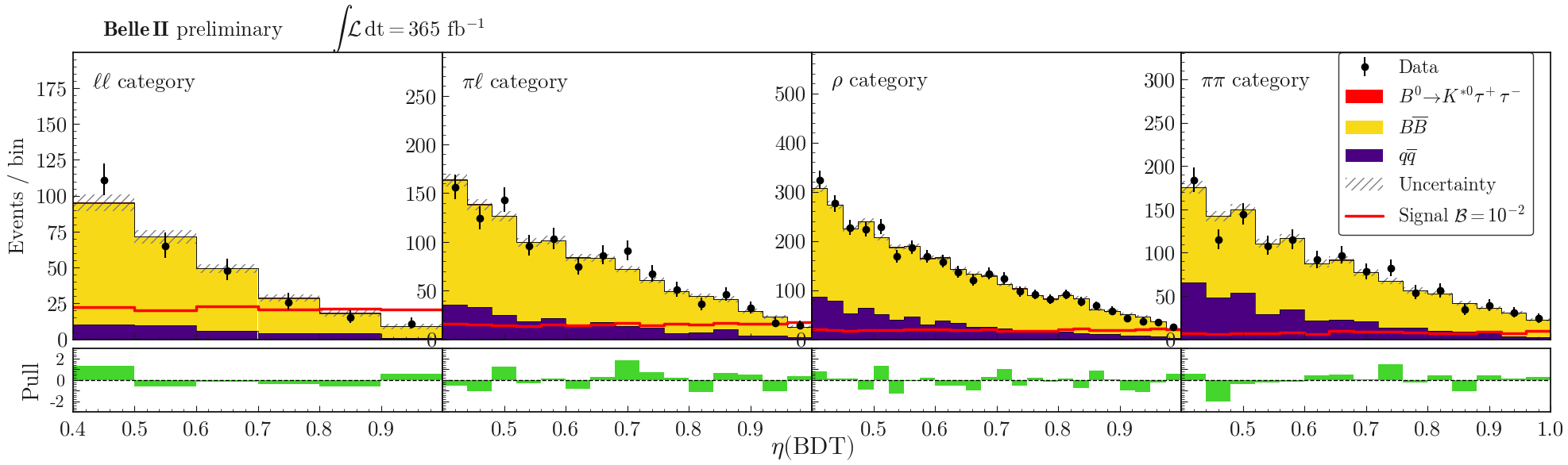}
\caption{
Distributions of \BDT in the SR for the four signal categories. 
The fit results are shown for the two background components (\BBbar and \qqbar) and the \BdToKsttautau signal, with a fitted \Branching of \BFdataShort. A \BdToKsttautau signal distribution, scaled assuming a \Branching of $10^{-2}$, is shown as reference. 
The bottom panel shows the pull distributions. 
}
\label{fig:distr_postfit}
\end{figure*}

The result of the fit to data is shown in Fig.~\ref{fig:distr_postfit}, corresponding to a measured \Branching of $\mathcal{B}(\BdToKsttautau)=\BFdata$. 
Compatibility between the data and fit result is assessed using simplified MC pseudo-experiments, and a \pvalue of \PVALdata is obtained. 
The measurement is statistically limited. 
The impact of the various systematic uncertainties on the \Branching is given in Table~\ref{tab:systematics}; the knowledge of the $B\to \D^{\ast\ast}\ell/\tau \nu$ decays \Branchings and the simulated sample size are the largest contributions. 
As no significant signal is observed, we obtain a 90\% C.L. upper limit of \ULdata. 

\begin{table}
	\caption{The systematic uncertainties for the \Branching of \BdToKsttautau, which were computed following the procedure in Ref.~\cite{Pinto:2023yob}. }
	\label{tab:systematics}
	\begin{ruledtabular}
	\begin{tabular}{lc}
		Source &  Impact on $\mathcal{B} \times 10^{-3}$   \\
		\hline
		$B\to D^{\ast\ast}\ell/\tau \nu$ \Branchings	&	0.29	\\
		Simulated sample size		    &	0.27	\\
		$q\bar q$ normalization                 &	0.18	\\
		ROE cluster multiplicity        &	0.17	\\
		$\pi$ and $K$ ID				&	0.14	\\
		$B$ decay \Branching			&	0.11	\\
		Combinatorial \BBbar normalization 		&	0.09	\\
		Signal and peaking \BzBzb normalization  &	0.07	\\
		Lepton ID						&	0.04	\\
		$\pi^0$ efficiency				&	0.03	\\
		\fzz							&	0.01	\\
        $N_{\FourS}$					&	0.01	\\
		$D \to \KL$ decays				&	0.01	\\
		Signal form factors				&	0.01	\\
		Luminosity						&	$< 0.01$	\\
		\hline
		Total systematics				&	0.52	\\
        \hline
		Statistics				        &	0.86	\\
	\end{tabular}
    \end{ruledtabular}
\end{table}

In summary, we present the first search for the \BdToKsttautau decays at \belletwo, utilizing the hadronic tagging technique. 
We analyze a \lumiyfours dataset collected by the \belletwo experiment at the SuperKEKB \epem collider. 
No evidence for a signal is observed, and an upper limit on the \Branching of \ULdata at the 90\% confidence level is set, assuming a signal with SM-like properties. 
This is the most stringent limit on the \BdToKsttautau decay to date. 

\begin{acknowledgments}
This work, based on data collected using the Belle II detector, which was built and commissioned prior to March 2019, was supported by
Higher Education and Science Committee of the Republic of Armenia Grant No.~23LCG-1C011;
Australian Research Council and Research Grants
No.~DP200101792, 
No.~DP210101900, 
No.~DP210102831, 
No.~DE220100462, 
No.~LE210100098, 
and
No.~LE230100085; 
Austrian Federal Ministry of Education, Science and Research,
Austrian Science Fund (FWF) Grants
DOI:~10.55776/P34529,
DOI:~10.55776/J4731,
DOI:~10.55776/J4625,
DOI:~10.55776/M3153,
and
DOI:~10.55776/PAT1836324,
and
Horizon 2020 ERC Starting Grant No.~947006 ``InterLeptons'';
Natural Sciences and Engineering Research Council of Canada, Compute Canada and CANARIE;
National Key R\&D Program of China under Contract No.~2022YFA1601903,
National Natural Science Foundation of China and Research Grants
No.~11575017,
No.~11761141009,
No.~11705209,
No.~11975076,
No.~12135005,
No.~12150004,
No.~12161141008,
No.~12475093,
and
No.~12175041,
and Shandong Provincial Natural Science Foundation Project~ZR2022JQ02;
the Czech Science Foundation Grant No.~22-18469S 
and
Charles University Grant Agency project No.~246122;
European Research Council, Seventh Framework PIEF-GA-2013-622527,
Horizon 2020 ERC-Advanced Grants No.~267104 and No.~884719,
Horizon 2020 ERC-Consolidator Grant No.~819127,
Horizon 2020 Marie Sklodowska-Curie Grant Agreement No.~700525 ``NIOBE''
and
No.~101026516,
and
Horizon 2020 Marie Sklodowska-Curie RISE project JENNIFER2 Grant Agreement No.~822070 (European grants);
L'Institut National de Physique Nucl\'{e}aire et de Physique des Particules (IN2P3) du CNRS
and
L'Agence Nationale de la Recherche (ANR) under Grant No.~ANR-21-CE31-0009 (France);
BMBF, DFG, HGF, MPG, and AvH Foundation (Germany);
Department of Atomic Energy under Project Identification No.~RTI 4002,
Department of Science and Technology,
and
UPES SEED funding programs
No.~UPES/R\&D-SEED-INFRA/17052023/01 and
No.~UPES/R\&D-SOE/20062022/06 (India);
Israel Science Foundation Grant No.~2476/17,
U.S.-Israel Binational Science Foundation Grant No.~2016113, and
Israel Ministry of Science Grant No.~3-16543;
Istituto Nazionale di Fisica Nucleare and the Research Grants BELLE2,
and
the ICSC – Centro Nazionale di Ricerca in High Performance Computing, Big Data and Quantum Computing, funded by European Union – NextGenerationEU;
Japan Society for the Promotion of Science, Grant-in-Aid for Scientific Research Grants
No.~16H03968,
No.~16H03993,
No.~16H06492,
No.~16K05323,
No.~17H01133,
No.~17H05405,
No.~18K03621,
No.~18H03710,
No.~18H05226,
No.~19H00682, 
No.~20H05850,
No.~20H05858,
No.~22H00144,
No.~22K14056,
No.~22K21347,
No.~23H05433,
No.~26220706,
and
No.~26400255,
and
the Ministry of Education, Culture, Sports, Science, and Technology (MEXT) of Japan;  
National Research Foundation (NRF) of Korea Grants
No.~2016R1-D1A1B-02012900,
No.~2018R1-A6A1A-06024970,
No.~2021R1-A6A1A-03043957,
No.~2021R1-F1A-1060423,
No.~2021R1-F1A-1064008,
No.~2022R1-A2C-1003993,
No.~2022R1-A2C-1092335,
No.~RS-2023-00208693,
No.~RS-2024-00354342
and
No.~RS-2022-00197659,
Radiation Science Research Institute,
Foreign Large-Size Research Facility Application Supporting project,
the Global Science Experimental Data Hub Center, the Korea Institute of
Science and Technology Information (K24L2M1C4)
and
KREONET/GLORIAD;
Universiti Malaya RU grant, Akademi Sains Malaysia, and Ministry of Education Malaysia;
Frontiers of Science Program Contracts
No.~FOINS-296,
No.~CB-221329,
No.~CB-236394,
No.~CB-254409,
and
No.~CB-180023, and SEP-CINVESTAV Research Grant No.~237 (Mexico);
the Polish Ministry of Science and Higher Education and the National Science Center;
the Ministry of Science and Higher Education of the Russian Federation
and
the HSE University Basic Research Program, Moscow;
University of Tabuk Research Grants
No.~S-0256-1438 and No.~S-0280-1439 (Saudi Arabia), and
Researchers Supporting Project number (RSPD2025R873), King Saud University, Riyadh,
Saudi Arabia;
Slovenian Research Agency and Research Grants
No.~J1-9124
and
No.~P1-0135;
Agencia Estatal de Investigacion, Spain
Grant No.~RYC2020-029875-I
and
Generalitat Valenciana, Spain
Grant No.~CIDEGENT/2018/020;
The Knut and Alice Wallenberg Foundation (Sweden), Contracts No.~2021.0174 and No.~2021.0299;
National Science and Technology Council,
and
Ministry of Education (Taiwan);
Thailand Center of Excellence in Physics;
TUBITAK ULAKBIM (Turkey);
National Research Foundation of Ukraine, Project No.~2020.02/0257,
and
Ministry of Education and Science of Ukraine;
the U.S. National Science Foundation and Research Grants
No.~PHY-1913789 
and
No.~PHY-2111604, 
and the U.S. Department of Energy and Research Awards
No.~DE-AC06-76RLO1830, 
No.~DE-SC0007983, 
No.~DE-SC0009824, 
No.~DE-SC0009973, 
No.~DE-SC0010007, 
No.~DE-SC0010073, 
No.~DE-SC0010118, 
No.~DE-SC0010504, 
No.~DE-SC0011784, 
No.~DE-SC0012704, 
No.~DE-SC0019230, 
No.~DE-SC0021274, 
No.~DE-SC0021616, 
No.~DE-SC0022350, 
No.~DE-SC0023470; 
and
the Vietnam Academy of Science and Technology (VAST) under Grants
No.~NVCC.05.12/22-23
and
No.~DL0000.02/24-25.

These acknowledgements are not to be interpreted as an endorsement of any statement made
by any of our institutes, funding agencies, governments, or their representatives.

We thank the SuperKEKB team for delivering high-luminosity collisions;
the KEK cryogenics group for the efficient operation of the detector solenoid magnet and IBBelle on site;
the KEK Computer Research Center for on-site computing support; the NII for SINET6 network support;
and the raw-data centers hosted by BNL, DESY, GridKa, IN2P3, INFN, 
and the University of Victoria.

\end{acknowledgments}

\ifthenelse{\boolean{wordcount}}%
{ \nobibliography{references} }
{ 
\bibliographystyle{apsrev4-2} 
\bibliography{references}

\begin{thebibliography}{37}
\bibitem [{cha()}]{chargeConj}%
  \BibitemOpen
  \href@noop {} {}\bibinfo {note} {Charge conjugate modes are implied
  throughout the text.}\BibitemShut {Stop}%
\end{thebibliography}
  
}

\end{document}